\documentclass{PoS}
\usepackage{epsfig}

\PoS{PoS(LAT2005)231}

\def\Oa{O($a$)}
\def\Oas{O($a^2$)}
\def\mpi2{m_\pi^2}

\newcommand{\be}{\begin{enumerate}}
\newcommand{\ee}{\end{enumerate}}
\newcommand{\bi}{\begin{itemize}}
\newcommand{\ei}{\end{itemize}}
\newcommand{\bflr}{\begin{flushright}}
\newcommand{\eflr}{\end{flushright}}
\newcommand{\bfll}{\begin{flushleft}}
\newcommand{\efll}{\end{flushleft}}
\newcommand{\ben}{\begin{enumerate}}
\newcommand{\een}{\end{enumerate}}
\newcommand{\beq}{\begin{equation}}
\newcommand{\eeq}{\end{equation}}
\newcommand{\beqn}{\begin{eqnarray}}
\newcommand{\eeqn}{\end{eqnarray}}
\newcommand{\nn}{\nonumber}

\def\xlf{\raisebox{+0.2em}{\boldmath{$\chi$}}\hspace{-0.2ex}\raisebox{-0.2em}{L}\hspace{-0.6ex}\raisebox{+0.12em}{F}\hspace{1mm}}

\title{Scaling test of quenched Wilson twisted mass QCD at maximal twist }

\ShortTitle{Scaling test of quenched Wilson twisted mass QCD at maximal twist }

\author{
\vspace*{0cm}\\
\hspace*{11.5cm}DESY~05-189\\
}

\author{K.~Jansen, \speaker{M.~Papinutto}, A.~Shindler, I.~Wetzorke\\
        John von Neumann-Institut f\"ur Computing NIC, Platanenallee 6, 
        15738 Zeuthen, Germany\\
        E-mail: \email{karl.jansen@desy.de}, \email{mauro.papinutto@desy.de}, 
        \email{andrea.shindler@desy.de}, \email{ines.wetzorke@desy.de}}

\author{C.~Urbach\\
        John von Neumann-Institut f\"ur Computing NIC, Platanenallee 6,
	15738 Zeuthen, Germany \&\\ 
        Institut f\"{u}r Theoretische Physik, Freie Universit\"{a}t Berlin,
        Arnimallee 14, 14195 Berlin, Germany\\
        E-mail: \email{carsten.urbach@desy.de}}

\abstract{We present the results of an extended scaling test of quenched
Wilson twisted mass QCD. We fix the twist angle by using two definitions
of the critical mass, the first obtained by requiring the vanishing of the
pseudoscalar meson mass $m_{\rm PS}$ for standard Wilson fermions and 
the second by requiring
restoration of parity at non-zero value of the twisted mass
$\mu$ and subsequently extrapolating to $\mu\rightarrow 0$. Depending on
the choice of the critical mass we simulate at values of
$\beta \in [5.7,6.45]$, for a range of pseudoscalar meson masses $250\ {\rm
  MeV} \lesssim m_{\rm PS} \lesssim 1\ {\rm GeV}$ and we perform the continuum
limit for the pseudoscalar meson decay constant $f_{\rm PS}$ and various hadron
masses (vector meson $m_V$, baryon octet $m_{\rm oct}$ and baryon decuplet
$m_{\rm dec}$) at 
fixed value of $r_0 m_{\rm PS}$. For both definitions of the critical
mass, lattice artifacts are consistent with \Oa\ improvement. However,
with the second definition, large \Oas\ discretization errors present at small
quark mass with the first definition are strongly suppressed. 
The results in the continuum limit are in very good agreement with those 
from the Alpha and CP-PACS Collaborations.}

\FullConference{XXIIIrd International Symposium on Lattice Field Theory\\
		 25-30 July 2005\\
		 Trinity College, Dublin, Ireland}

\begin{document}

\section{Introduction}
\vspace*{-0.2cm}
Twisted mass QCD (tmQCD), whose action reads 
\vspace*{-0.1cm}
\begin{equation}
S[U,\psi,\bar\psi] = a^4 \sum_x \bar\psi(x) ( D_W + m_0 + i \mu
\gamma_5\tau_3 ) \psi(x)\; ,
\label{tmaction}
\end{equation}

\vspace*{-0.1cm}
\noindent has been proposed as an alternative to Wilson QCD because it is
not affected by the problem of unphysical zero modes and can lead to 
simplifications of the operator mixing pattern~\cite{Frezzotti:2000nk}.
At maximal twist (i.e. by setting $m_0$ to its critical value $m_c$ up 
to \Oa, with $\mu$ now the bare quark mass) it has been 
shown~\cite{Frezzotti:2003ni} that 
parity even correlators (and thus energies and matrix elements) are 
automatically \Oa\ improved. Due to these properties, tmQCD is a
very interesting candidate for dynamical simulations 
at small quark masses~\cite{andrea}.
However, right at small quark masses $\mu \lesssim a$~\footnote{powers 
of $\Lambda_{\rm QCD}$ required to match physical dimensions are in the
following understood}, 
large lattice artifacts have been observed when using a
definition of $m_c$ obtained by requiring the vanishing of 
$m_{\rm PS}^2$ with standard Wilson fermions (in the following we will
call it $m_c^{\rm pion}$)~\cite{Bietenholz:2004wv}. This kind of lattice
artifacts are obviously affecting also dynamical simulations. An analysis
based on Wilson Chiral Perturbation Theory
(W$\chi$PT)~\cite{Aoki:2004ta} suggested a definition of 
$m_c$ suitable to reach quark masses $\mu \sim a^2$. This definition
is obtained by requiring the vanishing of the PCAC quark mass
\vspace*{-0.0cm}
\beq
m_{\rm PCAC}=\lim_{x_0\rightarrow \infty}\frac{\sum_{\mathbf x}\langle
\partial_0 A_0^a(x)\; P^a(0)\rangle}{2\sum_{\mathbf x}\langle 
P^a(x)P^a(0)\rangle}\,\nn
\label{mPCAC}
\eeq

\vspace*{-0.1cm}
\noindent where $P^a=\bar{\psi}\gamma_5\frac{\tau^a}{2}\psi$,
$A_\mu^a=\bar{\psi}\gamma_\mu\gamma_5\frac{\tau^a}{2}\psi$. 

An analysis {\it \`a la} Symanzik beyond \Oa\ shows the presence of
\Oas\ cutoff effects enhanced at small pion mass, the most dangerous 
of which are of order $(a/m_{\rm PS}^2)^{2k}$ $k \geq 1$ which signal the presence of possibly
large lattice artifacts in the $m_{\rm PS}^2\lesssim a$ 
regime~\cite{Frezzotti:2005gi} (see also~\cite{Sharpe:2005rq}). 
The result of this analysis is that
there are two ways of reducing these large lattice artifacts: fixing the
\Oa\ ambiguity in $m_c$ as proposed in Refs.~\cite{Aoki:2004ta} (we will call 
this definition $m^{\rm PCAC}_c$) or add the clover term with
non-perturbatively determined $c_{SW}$ coefficient (and the
corresponding value of $m_c$). We investigate here the first proposal 
(see Refs.~\cite{vittorio} for the second possibility).  

\vspace*{-0.25cm}
\section{Determination of $m^{\rm PCAC}_c$}
\vspace*{-0.1cm}

\begin{figure}[ht]
\vspace*{-0.8cm}\hspace*{-0.3cm}
\epsfig{figure=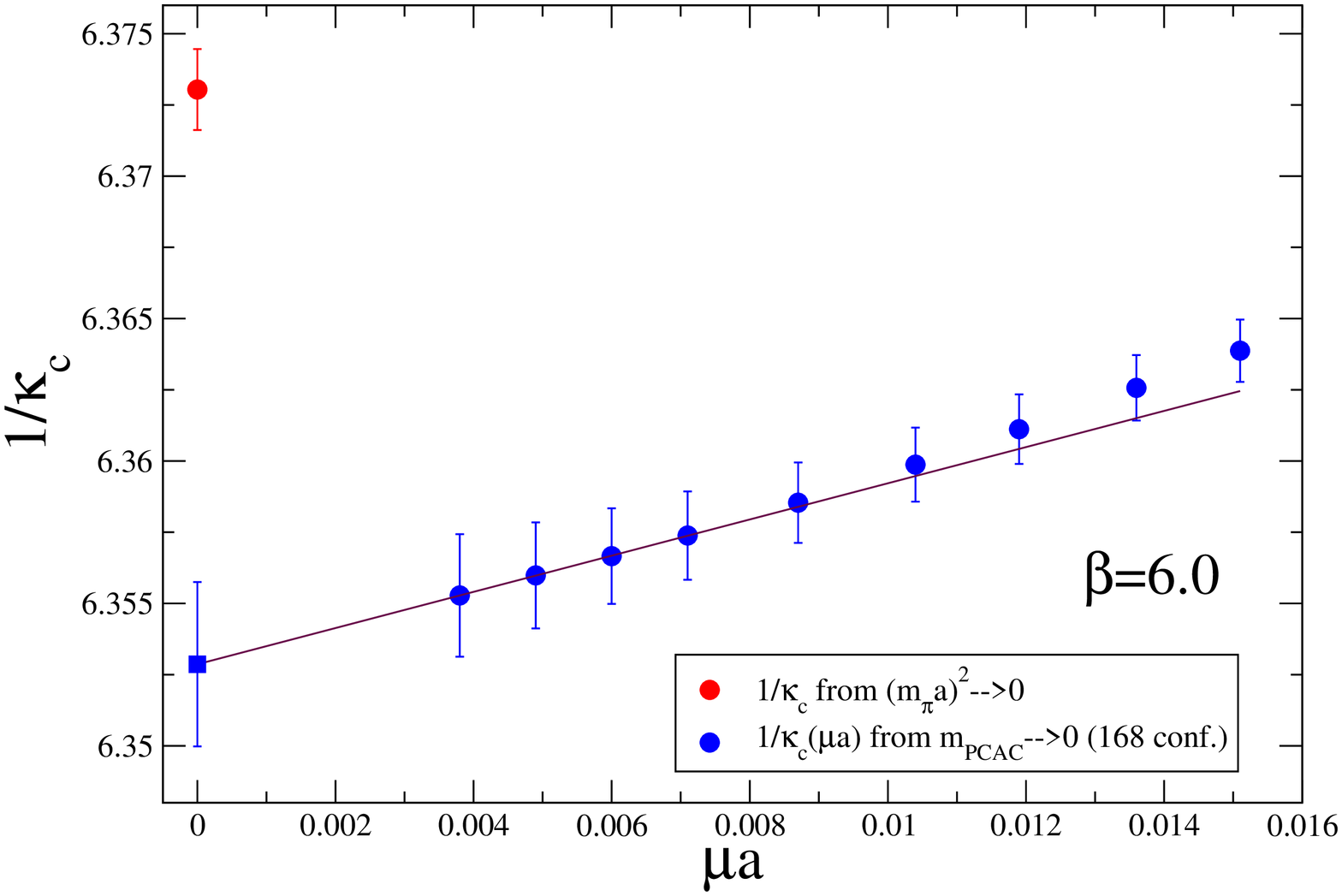,angle=270,width=0.55\linewidth}
\put(-170,-28){\epsfig{file=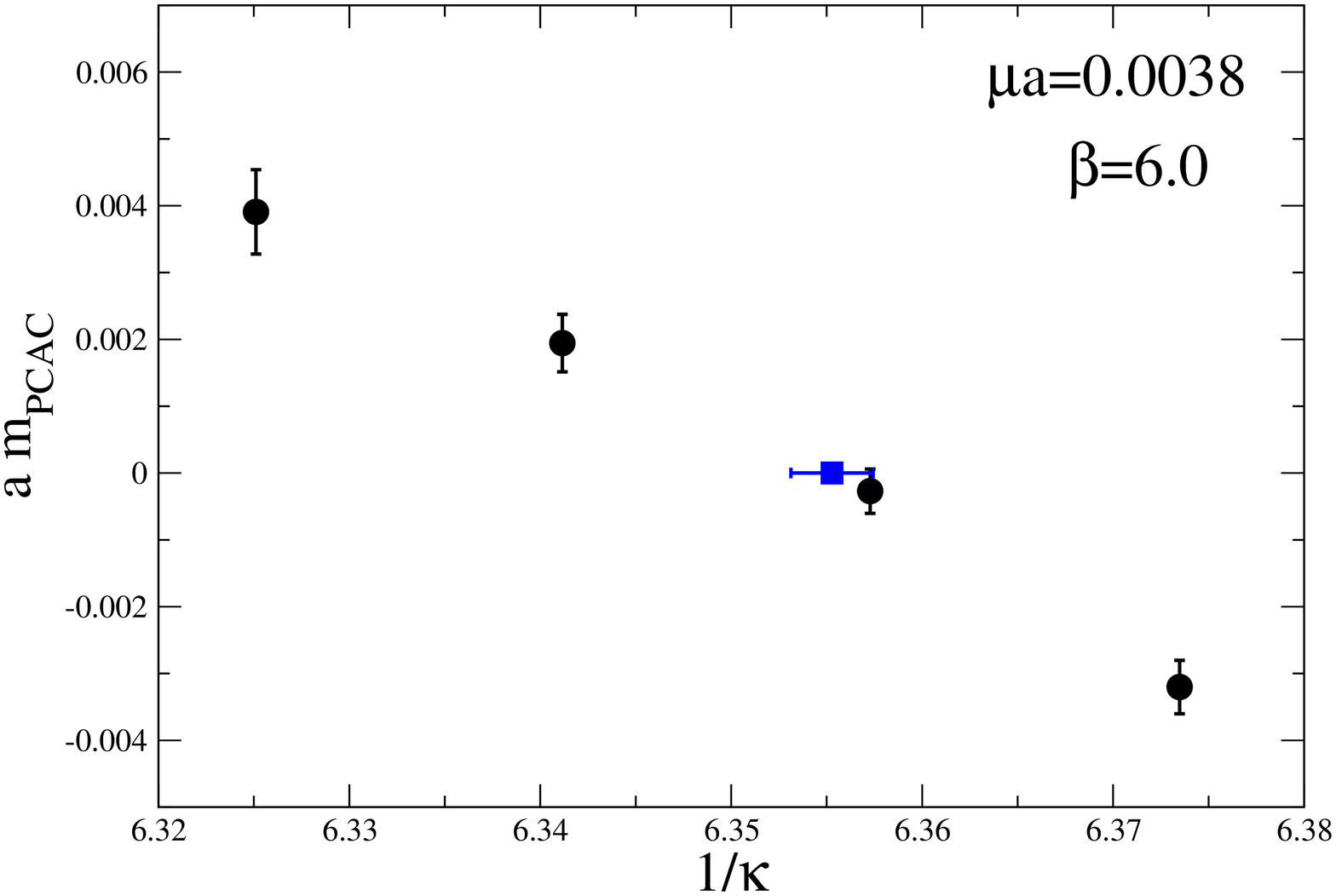,angle=270,width=0.20\linewidth}}
\put(-20,0){
\epsfig{figure=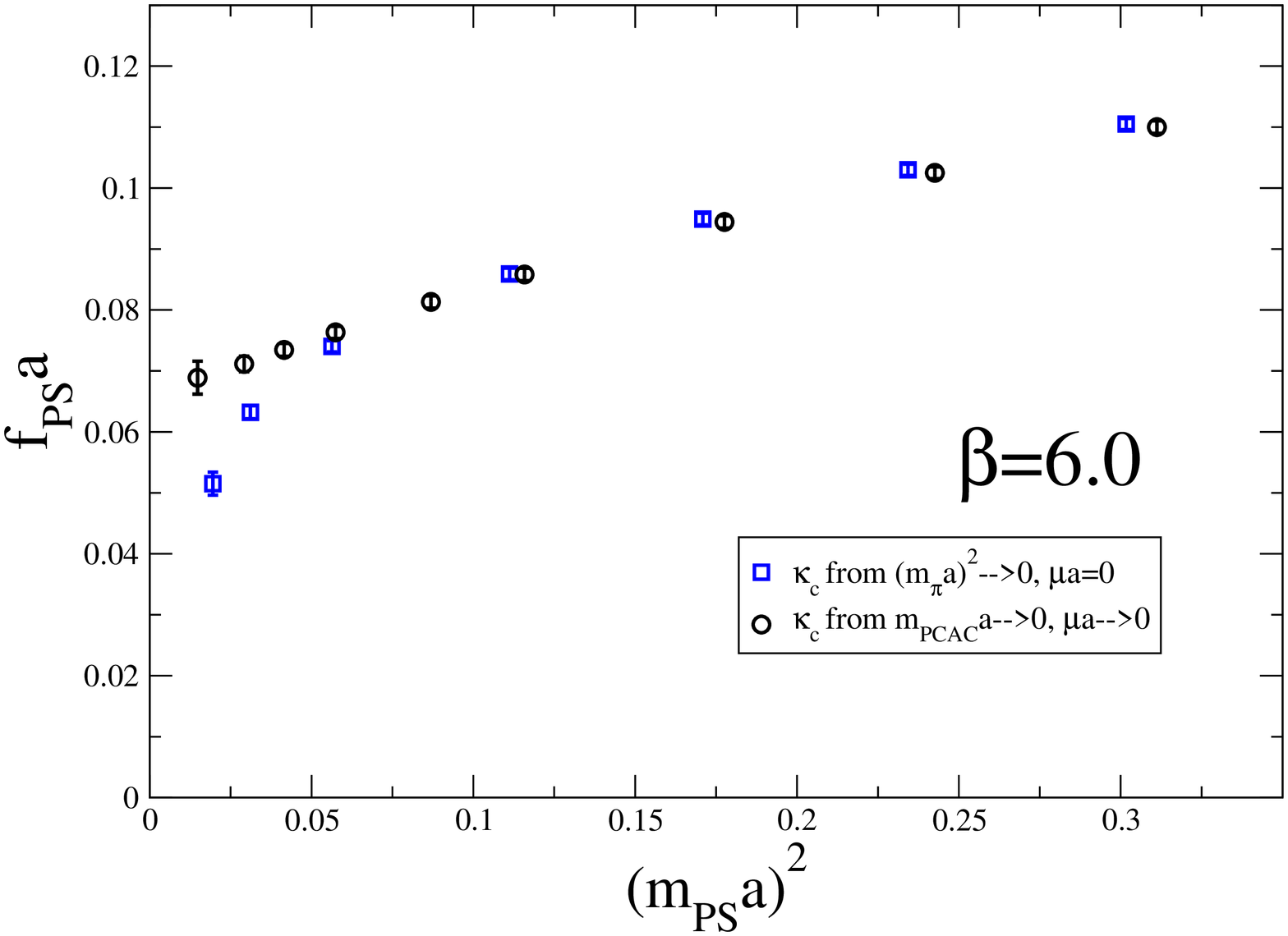,angle=270,width=0.55\linewidth}}
\vspace*{-0.5cm}
\caption{a. Determination of $m_c(\mu)$ ($\kappa_c^{-1}=2am_c+8$) 
for given values of $\mu$ and $\beta$ and
  extrapolation of $m_c(\mu)$ to $\mu=0$. b.
  $f_{\rm PS}$ as function of $m_{\rm PS}^2$ at $\beta=6.0$ 
  for the two definitions of
  $m_c$ ($m_c^{\rm pion}$ and $m_c^{\rm PCAC}$).}
\label{mpcac}
\end{figure}

$m_{\rm PCAC}$ depends, in the neighborhood of a given estimate of
$m_c$ (e.g. $m_c^{\rm pion}$), smoothly on both $m_0$ and $\mu$. 
Moreover, in the quenched approximation, a multiple mass solver can 
be used to compute the fermion propagator for different $\mu$ values at 
a given value of $m_0$. In view of these two facts, the procedure 
we adopted to determine $m^{\rm PCAC}_c$ is the 
following~\cite{Jansen:2005gf}:
\be
\item choose $n$ values of $\mu$ (with $\mu > a^2$) and $n'$ values 
of $m_0$ (in the vicinity of $m_c^{\rm pion}$) which cover 
the range where $m_{\rm PCAC}(\mu)$ is close to zero (in the present case $n=9$
and $n'=4$). 
\item find the values of $m_c(\mu)$ at which $m_{\rm PCAC}(\mu)$ is zero
  (see Fig.~\ref{mpcac}.a)
\item extrapolate $m_c(\mu)$ to $\mu=0$ (see
  Fig.~\ref{mpcac}.a)\footnote{in Ref.~\cite{Abdel-Rehim:2005gz} for each
  simulated value of $\mu$ the corresponding value of $m_c(\mu)$ were used.}. 
\ee 
In Ref.~\cite{Frezzotti:2005gi} one can find a theoretical
analysis that justifies this procedure and shows that the
value of $m^{\rm PCAC}_c$ in which we are interested can be 
obtained by a linear extrapolation from the region

\noindent $\mu > a^2$ down to $\mu=0$. 
$m_c^{\rm PCAC}$ has been determined for $\beta\in\{5.7,
5.85, 6.0, 6.2\}$ using statistics of O(100)-O(200) gauge configurations.

\vspace*{-0.25cm}
\section{Chiral behaviour at fixed $a$}
\vspace*{-0.1cm}

The effectiveness of $m_c^{\rm PCAC}$ in reducing the large lattice
artifacts observed at small quark mass is evident when considering the
chiral behaviour of two simple observables: $f_{\rm PS}$ and $m_{\rm PS}$.
$f_{\rm PS}$ can be extracted, without need of renormalization constants, 
by using the exact lattice PCVC relation 
$\langle\partial^*_\mu \tilde V^a_\mu(x) O(0)\rangle= -2 \mu 
\epsilon^{3ab}\langle P^b(x) 
O(0)\rangle$ (where $\partial^*_\mu$ is the lattice backward 
derivative and $\tilde V^a_\mu$ the point-splitted vector current).
It follows that, at maximal twist,   
\beq
f_{\rm PS}=\frac{2\mu}{m_{\rm PS}^2}
| \langle 0 | P^a | {PS}\rangle|\ .\nn 
\eeq
In Fig.~\ref{mpcac}.b one can compare the chiral behaviour of $f_{\rm
  PS}$ obtained with either $m_c^{\rm pion}$ or $m_c^{\rm PCAC}$. One
sees immediately that $m_c^{\rm PCAC}$ reduces the large lattice
artifacts present at small $\mu$ when using $m_c^{\rm pion}$ ($f_{\rm
  PS}$ is predicted to be linear in $m_{\rm PS}^2$ at one loop in
quenched $\chi$PT).

\begin{figure}[ht]
\vspace*{-0.8cm}\hspace*{-0.3cm}
\epsfig{figure=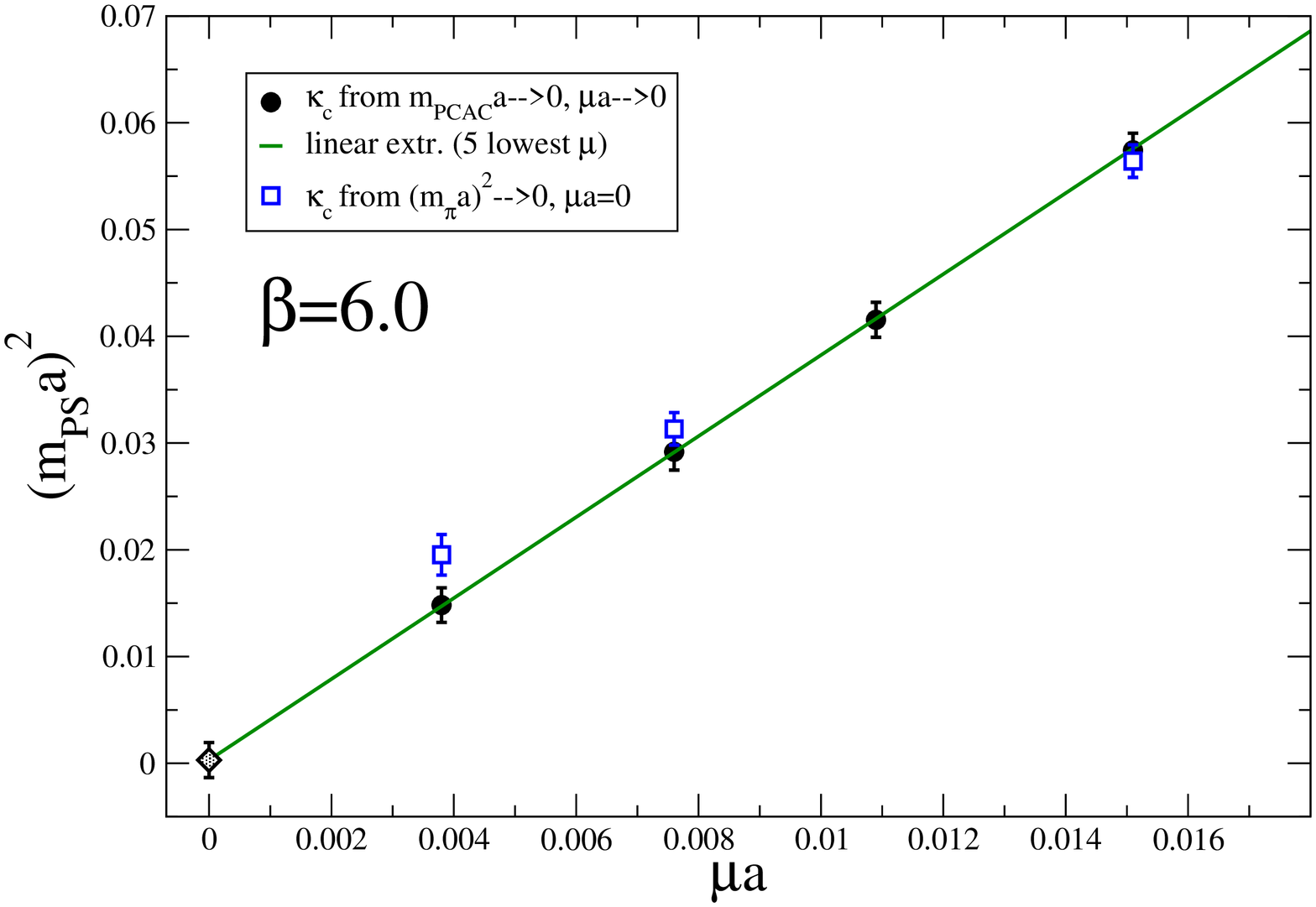,angle=270,width=0.55\linewidth}
\put(-20,0){
\epsfig{figure=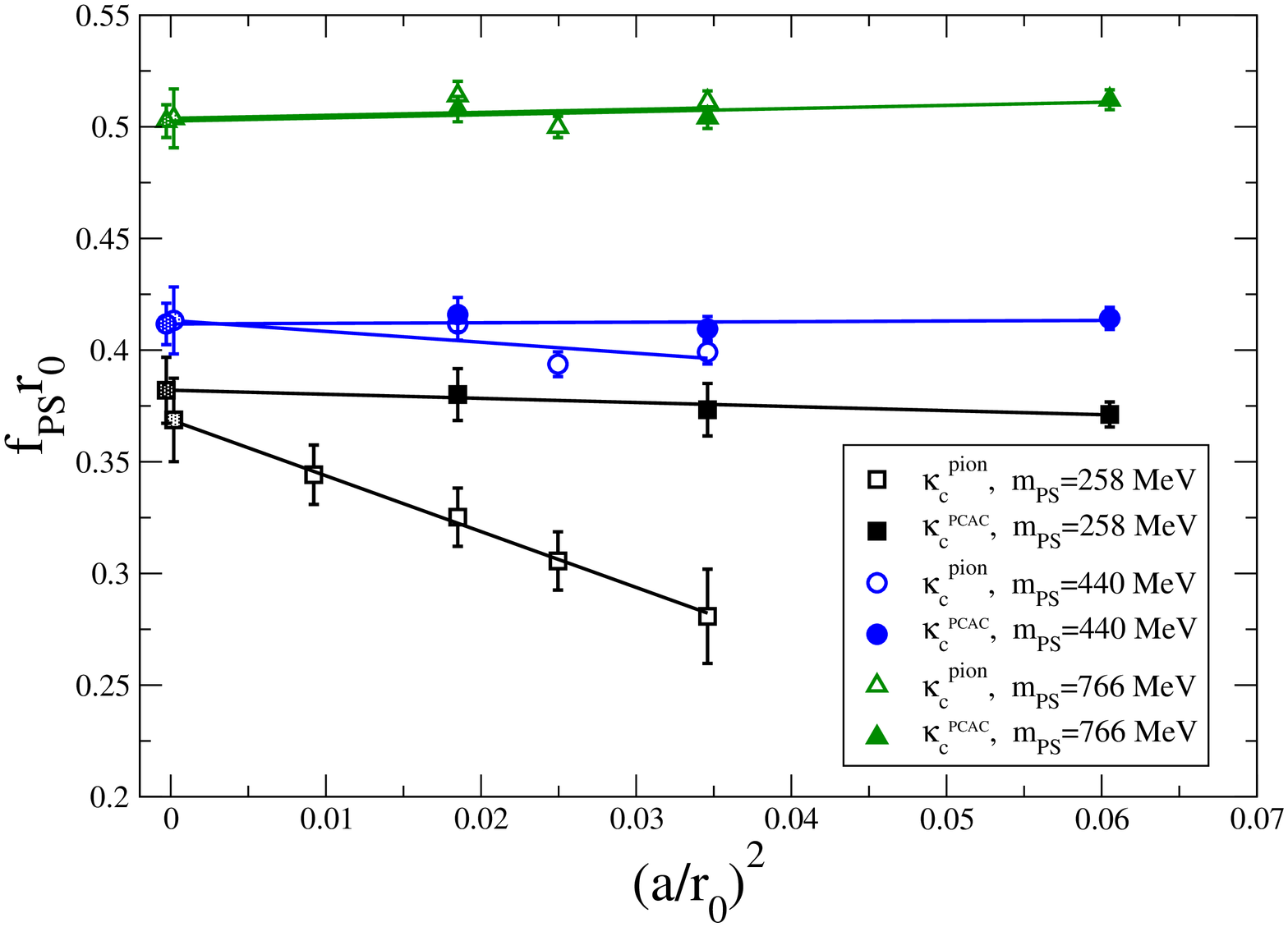,angle=270,width=0.55\linewidth}}
\vspace*{-0.5cm}
\caption{a. $m_{\rm PS}$ as function of $\mu$ at $\beta=6.0$ for
  both definitions $m_c^{\rm pion}$ and $m_c^{\rm
    PCAC}$. b. Scaling behaviour of $f_{\rm PS}r_0$ for 3 fixed values of 
$m_{\rm PS}r_0$. For each definition of $m_c$ an indipendent fit has been 
performed.}
\label{mpi}
\end{figure}

Using the integrated PCVC relation, it is also possible to 
prove~\cite{Frezzotti:2005gi} that, by using $m_c^{\rm PCAC}$ and in the 
region $\mu > a^2$,
$m_{PS}^2$ is linear with $\mu$ up to small $a^4$ 
cut-off effects~\footnote{chiral logs and other O($m_{\rm PS}^2$)
  contributions are here assumed to be negligible} 
(i.e. \Oas\ lattice artifacts are proportional to 
$\mu$). This is qualitatively confirmed by our data as shown 
in Fig.~\ref{mpi}.a. 

\vspace*{-0.25cm}
\section{Scaling behaviour}
\vspace*{-0.1cm}

We present now results concerning the scaling behaviour 
of $f_{\rm PS}r_0$, $m_Vr_0$, $m_{\rm oct}r_0$ and $m_{\rm dec}r_0$ at
fixed value of $m_{\rm PS}r_0$ (where $r_0$ is the Sommer scale).
More details concerning meson quantities can be found in 
Ref.~\cite{Jansen:2005kk}. 
$m_V$ has been extracted by using either the spatial  
component of the axial vector or the temporal component of the 
tensor as interpolating operators (in
the following we will quote only the latter, which systematically present
lower statistical errors). 
$m_{\rm oct}$ and $m_{\rm dec}$ have been extracted by using respectively
$\epsilon^{ABC}((d^A)^T C\gamma_5 u^B)u^C_\alpha$ and 
$\epsilon^{ABC}((u^A)^T C\gamma_k u^B)u^C_\alpha$ as interpolating
operators. 
The parameters of the simulations can be found in
Tab.~\ref{para}. We have simulated quark masses $\mu$ corresponding to 
$235 {\rm MeV} \leq m_{PS} \leq 1.0 {\rm GeV}$ (where the scale $r_0$, as 
will be explained below, has been fixed through the $\rho$ mass).  
The scaling behaviour of $f_{\rm PS}r_0$, as shown in 
Fig.~\ref{mpi}.b, is clearly linear in $(a/r_0)^2$. However, 
$m_c^{\rm pion}$ gives large \Oas\ effects at small masses,
effects that are drastically reduced by the use of $m_c^{\rm PCAC}$. 
This obviously influences the scaling region which starts at $\beta=6.0$ 
for $m_c^{\rm pion}$ and at $\beta=5.85$ 
for $m_c^{\rm PCAC}$. We perform indipendent continuum extrapolations
for the two choices of $m_c$ and the results are in good agreement. 
In the case of $m_c^{\rm pion}$, due to
the highest slope for the lowest quark mass, we needed an additional point 
at $\beta=6.45$ in order to control the extrapolation.  
$m_Vr_0$, $m_{\rm oct}r_0$ and $m_{\rm dec}r_0$ have 
been thus computed by using only $m_c^{\rm PCAC}$. 

\begin{figure}[b]
\vspace*{-0.7cm}\hspace*{-0.3cm}
\epsfig{figure=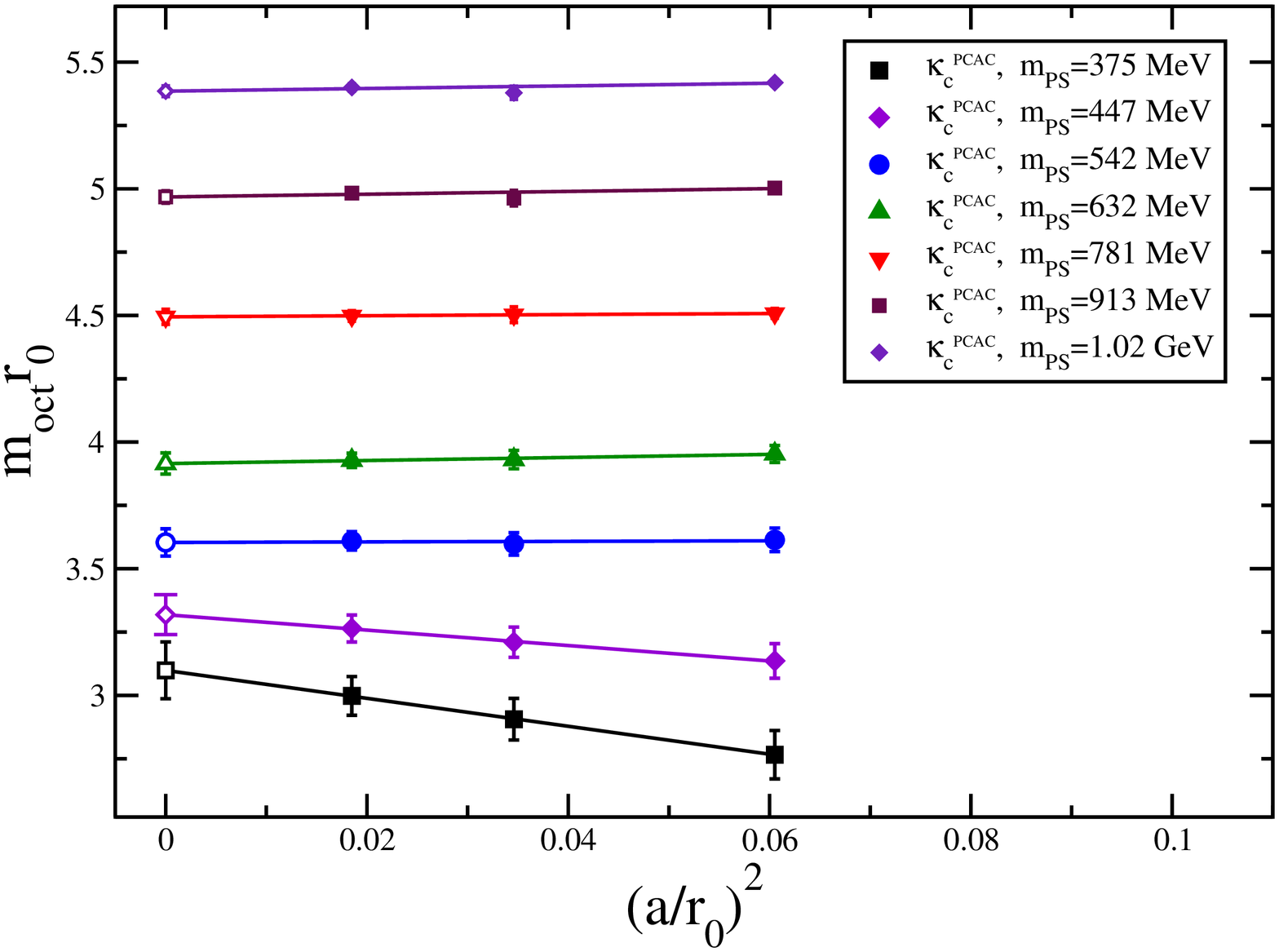,angle=270,width=0.55\linewidth}
\put(-20,0){
\epsfig{figure=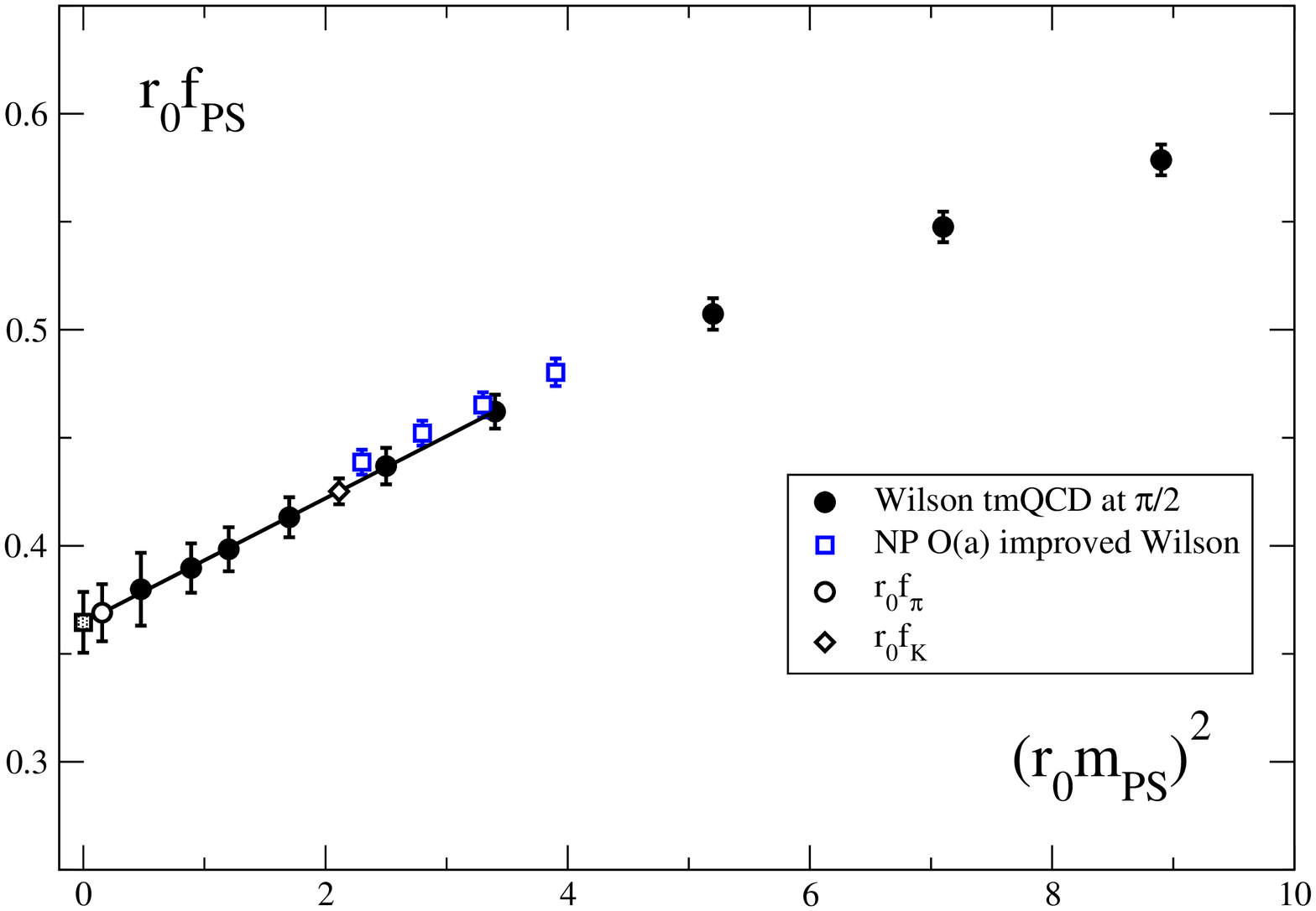,angle=270,width=0.55\linewidth}}
\vspace*{-0.5cm}
\caption{a. Scaling behaviour of $m_{\rm oct}r_0$ for 7 fixed values of 
$m_{\rm PS}r_0$ (only $m_c^{\rm PCAC}$ used). b. Continuum limit of 
$f_{\rm PS}r_0$ as function of $(m_{\rm PS}r_0)^2$ (only $m_c^{\rm
  PCAC}$ used). The empty squares are results taken from~\cite{Garden:1999fg}} 
\label{moct}
\end{figure}

Since we have simulated down to $m_{\rm PS}$ of 235 MeV, finite size 
effects (FSE)
can be quite relevant. In order to check for FSE we performed two 
additional simulations at $\beta=5.85$ on volumes of $12^3\times32$ and  
$14^3\times32$ in order to extend the results of 
Ref.~\cite{Guagnelli:2004ww} at smaller masses.
For meson quantities ($m_{\rm PS}$, $f_{\rm PS}$ and $m_V$) FSE are 
negligible for all the quark masses starting from the
third smallest one; on the two smallest masses they are in practice 
below the statistical accuracy of our data. For baryon masses, instead,
FSE are very large for the two smallest masses and still relevant on the
next three. Since the sensitivity required to study FSE on the smallest
two masses is computationally very expensive, we chose here to correct
only the data from the third value of $\mu$ on 
(corresponding to $m_{\rm PS}=375$ MeV). The results for the scaling
behaviour of $m_{\rm oct}r_0$ are shown in Fig.~\ref{moct}.a from which 
the lattice artifacts appear to be \Oas\ down to $\beta=5.85$ 
and of relatively small size.    
  
\begin{table}
\vspace*{-0.3cm}
{\footnotesize \begin{center}
\begin{tabular}{|c||c|c|c|c|c|c|}
\hline
\hline
$\beta$ & 5.70 &  5.85 & 6.00 & 6.10 & 6.20  & 6.45 \\
\hline   
$a$ (fm) & 0.171 & 0.123 & 0.093 & 0.079 & 0.068 & 0.048 \\
$L/a$ & 12 & 16 & 16 & 20 & 24 & 32\\
$T/a$ & 32 & 32 & 32 & 40 & 48 & 64\\
\hline
$N_{\rm conf}$ ($m_c^{\rm pion}$) & 600 & 378 & 387 & 300 & 260 & 182 \\
 \hline
$N_{\rm conf}$ ($m_c^{\rm PCAC}$) & 600 & 500 & 400 & & 300 & \\
\hline
\hline
\end{tabular}
\end{center}}
\vspace*{-0.3cm}
\caption{Parameters of the simulations}
\vspace*{-0.5cm}
\label{para}
\end{table}

\vspace*{-0.25cm}
\section{Continuum limit}
\vspace*{-0.1cm}

\begin{figure}[b]
\vspace*{-0.5cm}
\hspace*{-0.5cm}
\epsfig{figure=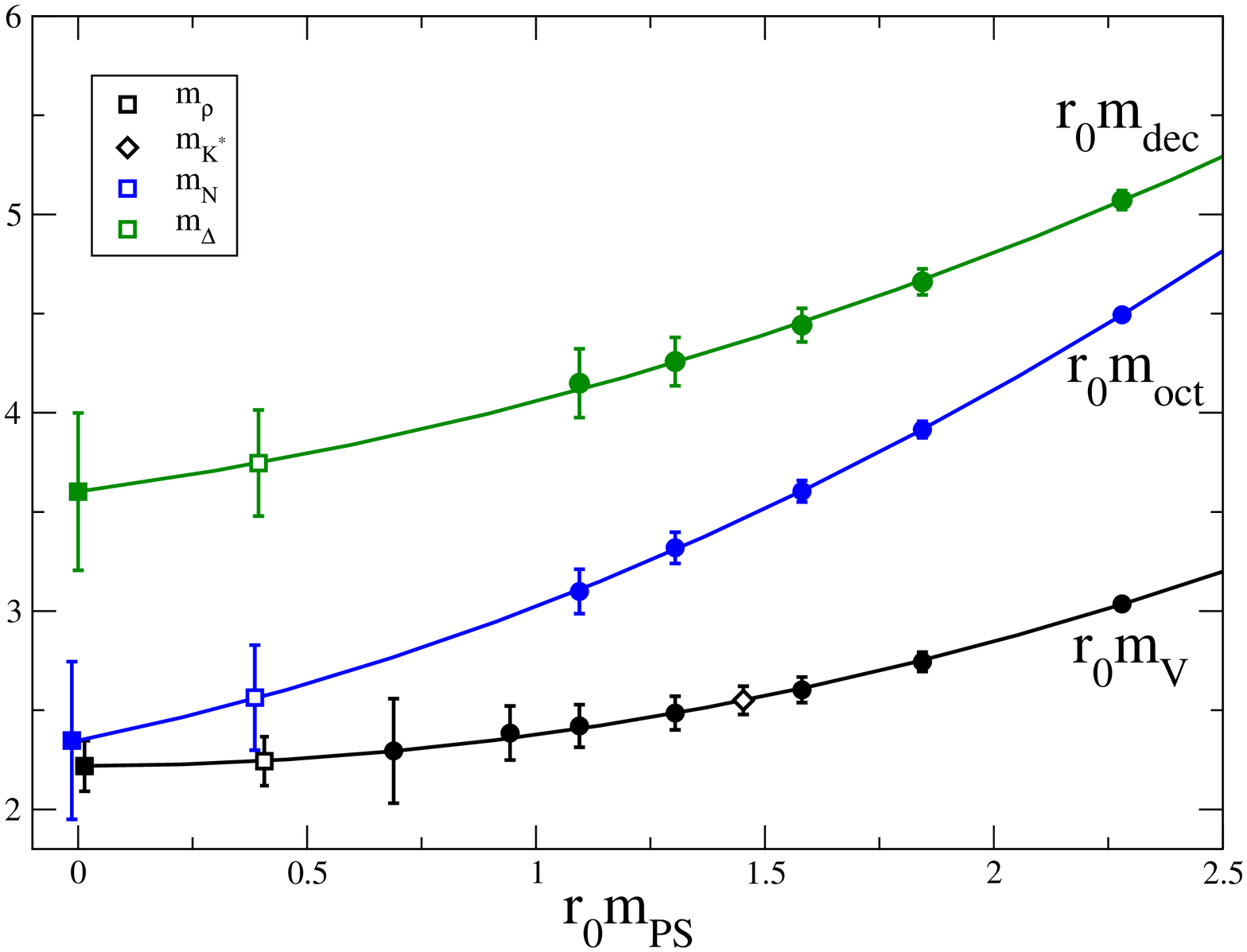,angle=270,width=0.55\linewidth}
\put(-20,0){
\epsfig{figure=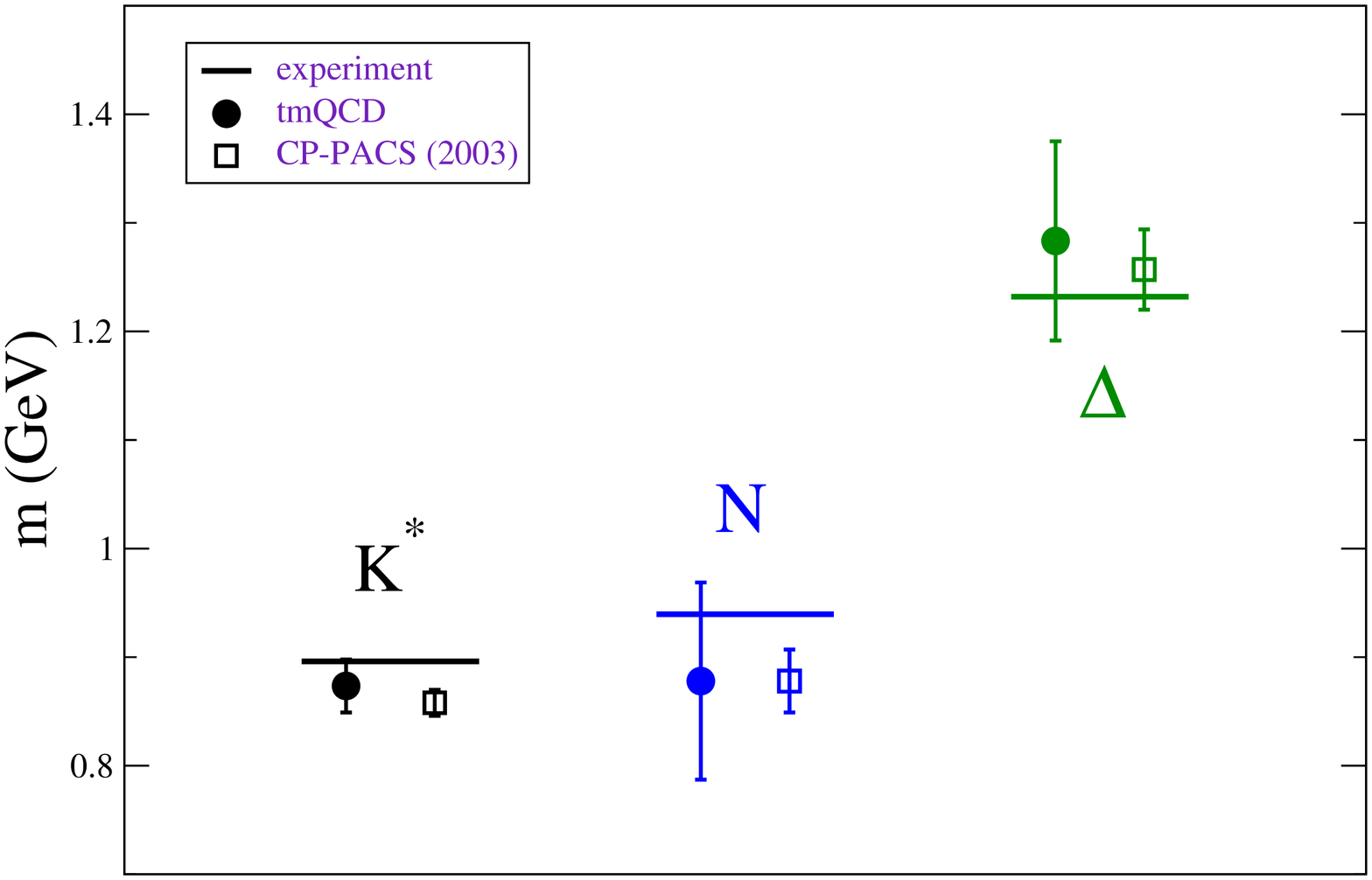,angle=270,width=0.55\linewidth}}
\vspace*{-0.5cm}
\caption{a. Continuum limit of $m_Vr_0$, $m_{\rm oct}r_0$, $m_{\rm dec}r_0$
as function of $m_{\rm PS}r_0$ (only $m_c^{\rm
  PCAC}$ used). b. Comparison of our results for $m_{K^*}$, $m_N$,
$m_\Delta$ with those of Ref.~\cite{Aoki:2002fd}.} 
\label{mhad}
\end{figure}

Results for the continuum limit of $f_{\rm PS}r_0$, $m_Vr_0$, $m_{\rm
  oct}r_0$ and $m_{\rm dec}r_0$ are presented in 
Fig.~\ref{moct}.b and~\ref{mhad}.a. Our determinations of $f_{\rm PS}$
and $m_V$ are in good agreement with those from the 
ALPHA Coll.~\cite{Garden:1999fg} with non-perturbatively \Oa\ improved Wilson
fermions (the comparison for $f_{\rm PS}$ is shown in Fig.~\ref{moct}.b).
For the chiral extrapolation we used the form $f_{PS}$, $m_V$
 $\sim A+B m_{PS}^2$ and $m_{\rm oct}$, $m_{\rm dec}$ 
$\sim A+B m_{PS}+C m_{PS}^2$. In order to compare with the results of 
the CP-PACS Coll.~\cite{Aoki:2002fd}, we fixed the scale $r_0$ through the
$\rho$ mass $m_\rho$, obtaining $r_0=0.576 fm$~\footnote{ 
had we fixed the scale through $f_K$ we would have obtained $r_0=0.508 fm$}. 
As a prediction we get $f_\pi$, $m_N$, $m_\Delta$ and
(working in the $SU(3)$ symmetric limit) $f_K$ and $m_{K^*}$. The
results can be found in Fig.~\ref{mhad}.b and Tab.~\ref{fpi} together
with those of Ref.~\cite{Aoki:2002fd} and turn out to be in very
good agreement. Notice however that in the present work quantities are
\Oa\ improved and pseudoscalar masses significantly
smaller than those in Ref.~\cite{Aoki:2002fd} (where standard Wilson
fermions were used) have been simulated.

\vspace*{-0.25cm}
\section{Conclusions and Outlook}
\vspace*{-0.1cm}

\begin{table}[ht]
\vspace*{-0.0cm}
{\footnotesize \begin{center}
\begin{tabular}{|c||c|c|c|}
\hline
\hline
 & $f_\pi$ (MeV) & $f_K$ (MeV) & $f_K/f_\pi$\\
\hline   
exp. & 132 & 160 & 1.22\\
tmQCD & 126(5) & 146(3) & 1.15(5)\\
CP-PACS & 120(6) & 139(5) & 1.16(3)\\
\hline
\hline
\end{tabular}
\end{center}}
\vspace*{-0.3cm}
\caption{Pseudoscalar meson decay constants from the present work and 
from Ref.~\cite{Aoki:2002fd} (in Ref.~\cite{Aoki:2002fd}, tadpole-improved
  perturbation theory has been used to renormalize the axial current).}
\vspace*{-0.0cm}
\label{fpi}
\end{table}

In the present study we have extrapolated to the continuum (taking 
into account possible FSE) meson
quantities ($f_{PS}$ and $m_V$) and baryon masses ($m_{\rm oct}$ and
$m_{\rm dec}$) down to pseudoscalar messon masses of 235 MeV and 375
MeV respectively.  
We have presented a strong evidence that lattice artifacts are \Oas\ for
both definitions of $m_c$ ($m_c^{\rm pion}$ and $m_c^{\rm PCAC}$) and
moreover that the use of $m_c^{\rm PCAC}$ drastically reduces the
chirally enhanced \Oas\ lattice artifacts present at small quark masses 
when using 
$m_c^{\rm pion}$. Our results for the continuum extrapolated quantities 
are in good agreement with those from the ALPHA~\cite{Garden:1999fg} 
and CP-PACS~\cite{Aoki:2002fd} Collaborations. This is very
encouraging in view of future dynamical simulations~\cite{andrea}. There
are however other aspects of tmQCD that it is worth to investigate,
for instance the problem of isospin breaking (see
Ref.~\cite{Jansen:2005cg} for an exploratory study in the quenched 
approximation). This problem is practically very important 
for phenomenological applications of tmQCD
and also strictly related to the phase structure of
tmQCD in the neighborhood of the critical 
point~\cite{Aoki:2004ta,Scorzato:2004da} and thus directly relevant for
dynamical symulations.

\end{document}